\documentclass[pre,twocolumn,amsmath,amssymb,floatfix]{revtex4}

\usepackage{graphicx}
\usepackage{dcolumn} 
\usepackage{bm}      
\usepackage{color}

\begin{document}

\title{Sign cancellation and scaling in the vertical component of 
       velocity and vorticity in rotating turbulence}
\author{E.~Horne$^{1}$ and P.D.~Mininni$^{2,3}$}
\affiliation{$^1$ Laboratoire de Physique de l'Ecole Normale Sup\'erieure 
de Lyon, 46 all\'ee d'Italie F-69007 Lyon, France. \\
             $^2$ Departamento de F\'\i sica, Facultad de Ciencias 
Exactas y Naturales, Universidad de Buenos Aires and IFIBA, CONICET, 
Ciudad Universitaria, 1428 Buenos Aires, Argentina. \\
             $^3$ NCAR, P.O. Box 3000, Boulder, Colorado 80307-3000, U.S.A.}
\date{\today}

\begin{abstract}
We study sign changes and scaling laws in the Cartesian components of the velocity and vorticity of rotating turbulence, in the helicity, and in the components of vertically-averaged fields. Data for the analysis is provided by high-resolution direct numerical simulations of rotating turbulence with different forcing functions, with up to $1536^3$ grid points, with Reynolds numbers between $\approx 1100$ and $\approx 5100$, and with moderate Rossby numbers between $\approx 0.06$ and $\approx 8$. When rotation is negligible, all Cartesian components of the velocity show similar scaling, in agreement with the expected isotropy of the flow. However, in the presence of rotation only the vertical components of the fields show clear scaling laws, with evidence of possible sign singularity in the limit of infinite Reynolds number. Horizontal components of the velocity are smooth and do not display rapid fluctuations for arbitrarily small scales. The vertical velocity and vorticity, as well as the vertically-averaged vertical velocity and vorticity, show the same scaling within error bars, in agreement with theories that predict that these quantities have the same dynamical equation for very strong rotation.
\end{abstract}
\maketitle

\section{\label{sec:Intro}Introduction}

The dynamics of incompressible fluids is described by the Navier-Stokes equation, and is governed by competing and interacting processes such as nonlinear interactions, internal friction, external forces, and boundary conditions. The flow behavior is as diverse as the different ways these natural forces and processes can be combined.

However, when nonlinear interactions are sufficiently strong, the flow becomes turbulent, and certain statistical properties of the flow are believed to become universal. The turbulent regime, resulting from the predominance of nonlinear interactions over viscous dissipation, is present in numerous flows in nature, such as in geophysical and astrophysical flows. The ratio of the amplitude of these two processes is described through the Reynolds number $Re$, which can take values ​​as large as $Re \approx 10^8$ or higher in the atmosphere and in the oceans \cite{Pedlosky}, and $Re \approx 10^{12}$ or higher in astrophysics \cite{Brandenburg11}. While turbulence is often associated with very complicated and disordered flows, that is the case only for isotropic and homogeneous turbulence. When external forces such as rotation or stratification are present, the flow becomes highly anisotropic, and self-organization processes can take place in which this disorder can coexist with the development of ordered large-scale and long-living structures \cite{Cambon89,Waleffe93}.

An important example of anisotropic flows is given by rotating flows \cite{Cambon89,Cambon97,Cambon04}. Large-scale flows in the atmosphere and oceans are predominantly affected by the rotation of the earth \cite{Pedlosky}. Rotation is also important in many engineering flows \cite{Chen05}. The breaking of isotropy in a flow through rotation results in a quasi-two dimensional behavior for the velocity and vorticity fields, with the formation of large-scale columns in the velocity field \cite{Cambon89,Waleffe93,Cambon97}.

In rotating turbulent flows, quadratic quantities in the fields such as the energy, the helicity, and the enstrophy, are often characterized with isotropic and anisotropic spectra \cite{Cambon97,Bellet06,Mininni12} (or, equivalently, with second order structure functions \cite{Lamriben11,Mininni10}) following power laws in the inertial range. However, turbulent flows tend also to be intermittent \cite{Muller07,Mininni09}. Intermittency is caused by the presence of structures in the flow at different scales, highly localized in space and time. The proximity or remoteness of such structures can lead to rapid changes in the field derivatives, as well as to rapid changes in sign.

Characterization of intermittency is often done by studying probability density functions (PDFs) of velocity increments, and high order structure functions. While the second order moments of the PDFs (or the second order structure function) are related with the energy spectrum, in an intermittent flow higher order moments cannot be trivially inferred from the knowledge of the energy scaling. In simple terms, as the scale of interest is decreased, turbulent flows are increasingly more likely to develop strong gradients in the fields. This increase in the probability of extreme events with decreasing scale results in a breakdown of perfect scale invariance, in the development of non-Gaussian statistics, and in the need of more than one coefficient to characterize all moments of the PDF.

The characterization of these extreme events is an important part of the study of turbulence. While there are many tools to characterize intermittency in isotropic and homogeneous flows, its study in anisotropic flows is less developed. Numerical simulations and experiments indicate rotating turbulence is less intermittent than isotropic and homogeneous turbulence \cite{Baroud02,Baroud03,Muller07,Seiwert08,Mininni12}. However, for rotating flows, it has been argued that the observed scaling laws may actually be spurious, and the result of applying methods devised for isotropic turbulence to anisotropic flows \cite{Sagaut}. Moreover, while for strong rotation some theories predict decoupling between two-dimensional (2D) and three-dimensional (3D) modes \cite{Galtier03,Chen05,Bourouiba08,Teitelbaum12} (although other authors claim that the modes never decouple in infinite domains \cite{Cambon04}), and that the vertical velocity and vorticity behave as passive scalars \cite{Embid96,Babin96,Chen05}, the actual degree of decoupling and the scaling of individual components of the fields is hard to quantify.

In this work we use the cancellation exponent \cite{Ott92,Vainshtein94} to study the isotropic and anisotropic scaling of different quantities in rotating turbulence. The exponent gives information about the rapid changes in the sign of scalar quantities, and has been used before to characterize fluctuations of velocity and magnetic fields components in hydrodynamic turbulence and magnetohydrodynamic dynamos \cite{Ott92}, of the current density in 2D magnetohydrodynamic turbulence \cite{Sorriso02,Graham05}, of magnetic helicity in solar wind observations \cite{Bruno97}, and of helicity in isotropic and homogeneous hydrodynamic turbulence \cite{Imazio10}. We analyze data from high resolution direct numerical simulations (up to $1536^3$ grid points) and compute the cancellation exponent for the Cartesian components of the velocity and vorticity fields and for the helicity. Considering the symmetries of rotating flows and the strong anisotropy that develops, we also compute the cancellation exponent for the vertically-averaged velocity, vorticity, and helicity. We find that for strong rotation only the vertical component of the velocity and the vorticity show clear power law scaling, an indication of sign singularity for infinite Reynolds number. Moreover, the vertical velocity and vorticity in many of the simulations show the same scaling, in agreement with theories that predict that for strong rotation both quantities follow the same dynamics. The horizontal components of the fields are smoother and do not show strong sign fluctuations at small scales.  

\section{Rotating flows and numerical simulations}

\begin{table}
\caption{\label{table:runs}Parameters used in the simulations. $N$ is 
         the linear grid resolution, ${\bf f}$ the forcing [either Taylor 
         Green (TG) or Beltrami (ABC)], $k_F$ the forcing wavenumber, 
         $\nu$ is the viscosity, $\Omega$ is the rotation frequency, 
         $Re$ the Reynolds number, and $Ro$ the Rossby number. Runs ``T'' 
         have no net helicity, while runs ``A'' have maximal helicity 
         injection.}
\begin{ruledtabular}
\begin{tabular}{cccccccc}
Run &$N$  &${\bf f}$&$k_f$&     $\nu$        &$\Omega$ &$Re$ & $Ro$  \\
\hline
T1  &512  & TG      & 4   &$8\times10^{-4}$  & $0.4$   &1100  &$1.40$ \\ 
T2  &512  & TG      & 4   &$8\times10^{-4}$  & $1.6$   &1100  &$0.35$ \\
T3  &512  & TG      & 4   &$8\times10^{-4}$  & $8.0$   &1100  &$0.07$ \\
\hline
A1  &512  &ABC      & 7-8 &$6.5\times10^{-4}$& $0.06$  &1200  &$7.90$ \\
A2  &512  &ABC      & 7-8 &$6.5\times10^{-4}$& $7.0$   &1200  &$0.07$ \\
A3  &1536 &ABC      & 7-8 &$1.6\times10^{-4}$& $9.0$   &5100  &$0.06$ \\ 
\end{tabular}
\end{ruledtabular}
\end{table}

Before introducing the cancellation exponent, we briefly present in this section some results for rotating turbulence that motivate decisions in the way the numerical data is analyzed, and are also useful to interpret the results. For details of rotating turbulence, the reader is referred to \cite{Cambon04,Sagaut} and references therein. We also describe in this section the direct numerical simulations that were used for the analysis.

Incompressible rotating turbulence is described by the Navier-Stokes equations in a rotating frame, which for the velocity field ${\bf u}$ can be written as
\begin{equation}
\frac{\partial {\bf u}}{\partial t} + \mbox{\boldmath $\omega$} \times
    {\bf u} + 2 \mbox{\boldmath $\Omega$} \times {\bf u}  =
    - \nabla {\cal P} + \nu \nabla^2 {\bf u} + {\bf F} ,
\label{eq:momentum}
\end{equation}
and
\begin{equation}
\nabla \cdot {\bf u} =0 ,
\label{eq:incompressible}
\end{equation}
where $\mbox{\boldmath $\omega$} = \nabla \times {\bf u}$ is the vorticity, ${\cal P}$ is the total pressure modified by the centrifugal term and divided by the fluid mass density, and $\nu$ is the kinematic viscosity. The external force ${\bf F}$ drives the turbulence, and in the following the rotation axis is chosen in the $z$ direction, $\mbox{\boldmath $\Omega$} = \Omega \hat{z}$, with $\Omega$ the rotation frequency.

In the linearized case these equations accept helical waves as solutions, which correspond to inertial waves (see, e.g., \cite{Waleffe93}), and have dispersion relation $\omega=\pm 2 \mbox{\boldmath $\Omega$} \cdot {\bf k}/k$. In the nonlinear case and in wave turbulence theory, modes in Fourier space can thus be separated between 2D modes (with zero frequency, and therefore often called ``slow'' modes) and 3D modes (often called ``fast'' modes). The velocity associated with the slow modes can be obtained from a vertical average (see, e.g., \cite{Chen05}),
\begin{equation}
\overline{\bf u}(x,y) = \frac{1}{L} \int_0^L {\bf u}(x,y,z) dz,
\end{equation}
with $L$ the vertical size of the box. A vertically-averaged vorticity, which will be of interest for reasons explained below, can be computed in the same way,
\begin{equation}
\overline{\mbox{\boldmath $\omega$}}(x,y) = \frac{1}{L} \int_0^L 
    \mbox{\boldmath $\omega$}(x,y,z) dz.
\end{equation}
We can further decompose these vertically-averaged fields into a vector field in the $(x,y)$ plane (i.e., perpendicular to the rotation axis), and a vertical component parallel to the rotation axis. For the velocity field this results in
\begin{equation}
\overline{\bf u}(x,y) = \overline{\bf u}_\perp(x,y) + 
    \overline{u}_z(x,y) \hat{z} .
\end{equation}
The remaining of the velocity field (with spatial dependence in the vertical direction) is fully 3D and thus corresponds to ``fast'' modes. The same decomposition can be used for the vertically-averaged vorticity
\begin{equation}
\overline{\mbox{\boldmath $\omega$}}(x,y) = 
    \overline{\mbox{\boldmath $\omega$}}_\perp(x,y) + 
    \overline{\omega}_z(x,y) \hat{z} .
\end{equation}

In rotating turbulent flows, slow and fast modes interact through resonant and non-resonant triadic interactions. In wave turbulence theory only resonant interactions are considered to the lowest order in an expansion in terms of the Rossby number (assumed small). This results in a decoupling of the 2D modes in the limit of rapid rotation \cite{Waleffe93,Galtier03} (see however \cite{Cambon04} for the case of infinite domains). As a result, $\overline{\bf u}_\perp$ is expected to satisfy the 2D Navier-Stokes equation
\begin{equation}
\frac{\partial \overline{\bf u}_\perp}{\partial t} + 
    \overline{\bf u}_\perp \cdot \nabla \overline{\bf u}_\perp = 
    -\nabla \overline{\cal P} + \nu \nabla^2 \overline{\bf u}_\perp .
\label{eq:u2D}
\end{equation}
If decoupling of 2D and 3D modes is assumed, it also results that the equation for the vertically-averaged vertical velocity is
\begin{equation}
\frac{\partial \overline{u}_z}{\partial t} + 
    \overline{\bf u}_\perp \cdot \nabla \overline{u}_z = \nu \nabla^2 \overline{u}_z ,
\label{eq:vz}
\end{equation}
which tells us that the vertically-averaged vertical velocity is advected and diffused by $\overline{\bf u}_\perp$ as a passive scalar.

Taking the curl of Eq.~(\ref{eq:u2D}), we obtain the equation for the vertically-averaged vertical component of the vorticity,
\begin{equation}
\frac{\partial \overline{\omega}_z}{\partial t} + 
    \overline{\bf u}_\perp \cdot \nabla \overline{\omega}_z = 
    \nu \nabla^2 \overline{\omega}_z .
\label{eq:wz}
\end{equation}
This equation is again the equation of a 2D passive scalar, and therefore $\overline{\omega}_z$ should be also passively advected and diffused by $\overline{\bf u}_\perp$. In other words, both $\overline{u}_z$ and $\overline{\omega}_z$ follow the same equation under these approximations.

It should be noted that near-resonant interactions and higher-order resonances may break the decoupling, and a reduced model of rotating turbulence may be more complex than just 2D Navier-Stokes (as a matter of fact, the behavior of 2D modes in rotating turbulence is known to display differences with 2D turbulence, see e.g., \cite{Smith96,Smith05,Sen12,Teitelbaum12}, and asymptotic expansions also indicate that some coupling persists between 2D and 3D modes \cite{Julien06}). However, it is interesting to know whether the fact that at the level of resonant triads Eqs.~(\ref{eq:vz}) and (\ref{eq:wz}) are the same results in similar scaling for $\overline{u}_z$ and $\overline{\omega}_z$.

Another quantity of interest in incompressible turbulent flows is the helicity density,
\begin{equation}
h (x,y,z) = {\bf u} \cdot \mbox{\boldmath $\omega$} ,
\end{equation}
a scalar quantity that when integrated over volume ($H=\int {\bf u} \cdot \mbox{\boldmath $\omega$} dV$) is conserved in the limit of infinite Reynolds number \cite{Moffatt69}. As a result, in the ideal case helicity is a scalar that far from walls and on the average can only be transported by the flow \cite{Matthaeus08}. In forced turbulent flows, helicity has a direct cascade \cite{Andre77,Borue97,Chen03,Chen03b}, and in rotating turbulence helicity is known to affect the scaling of the energy spectrum \cite{Mininni09b}. Considering the symmetries of rotating flows, and in analogy with the vertically averaged velocity and vorticity introduced above, besides $h(x,y,z)$ we will also consider here a vertically-averaged helicity density
\begin{equation}
\overline{h} (x,y) = \frac{1}{L} \int_0^L h(x,y,z) dz .
\end{equation}

In the following sections, we compute cancellation exponents for the Cartesian components of the velocity ${\bf u}$ and vorticity $\mbox{\boldmath $\omega$}$, of the components of the vertically-averaged fields $\overline{\bf u}$ and $\overline{\mbox{\boldmath $\omega$}}$, and of the helicity densities $h$ and $\overline{h}$. The data for the analysis stems from direct numerical simulations of rotating turbulent flows in a 3D periodic domain of size $L=2\pi$. Equations (\ref{eq:momentum}) and (\ref{eq:incompressible}) are solved using a pseudospectral method, and evolved in time with a second order Runge-Kutta scheme. Six simulations were used, with spatial resolutions ranging from $512^3$ to $1536^3$ grid points, with at least three snapshots of the fields in the turbulent steady state of each run, and with two different types of forcing mechanisms: with Arn'old-Beltrami-Childress (ABC) forcing (fully helical, \cite{Mininni09b}), or with Taylor-Green (TG) forcing (non-helical, \cite{Mininni09}). The runs are described in detail in \cite{Mininni09,Mininni09b,Mininni10b}. Table \ref{table:runs} lists all the runs used in the analysis, and the parameters of each run: the forcing function used, the wavenumber $k_F$ at which the forcing was applied, the linear resolution $N$, and the Reynolds and Rossby numbers, defined respectively as
\begin{equation}
Re = \frac{UL}{\nu} , \,\,\,\,\, Ro = \frac{U}{2L \Omega} ,
\end{equation}
where $L=2\pi/k_F$ is the forcing scale, and $U$ is the r.m.s.~velocity.

\section{Cancellation exponent}

In a turbulent flow, intermittency is created by the presence of structures localized in space and in time. The proximity or remoteness of such structures can lead to rapid changes in the derivatives of the field, and also to rapid changes in sign. To study the scaling of such variations, the cancellation exponent was introduced in \cite{Ott92}. It is based on the definition of a signed measure, which is similar to a probability measure, only that it can take positive and negative values. 

Given a scalar quantity $f({\bf x})$ in a domain $Q(L)$ (where $L$ is the linear size of the total domain), the signed measure for a subdomain $Q_i(l)$ of linear size $l$ is defined as
\begin{equation}
\mu_{i}(l)=\frac{\int_{Q_{i}(l)} f(x) d^{3}x}{\int_{Q(L)} |f(x)| d^{3}x} .
\label{eq:mu}
\end{equation}
Here, the subdomain $Q_i(l)$ is defined such that a set $\{Q_i(l); \, i=1,2,\dots\}$ covers the entire domain $Q(L)$ without overlaps between subdomains, and $i$ is an index that labels the different subdomains. It follows from Eq.~(\ref{eq:mu}) that $-1 \le \mu_{i}(l) \le 1$. We can thus interpret $\mu_{i}(l)$ as the difference between the probability measure of the positive component of $f({\bf x})$ and the negative component of $f({\bf x})$.

For each scale $l$, we can now define a partition function by summing over all the subdomains with size $l$ that cover the entire domain,
\begin{equation}
\chi(l)=\sum_i |\mu_{i}(l)| .
\label{eq:chi}
\end{equation}
For a usual (unsigned) probability measure, $\chi(l)=1$; that value is also obtained for $\mu_{i}(l)$ for sufficiently small $l$ if the function $f({\bf x})$ is smooth. However, when there are sign cancellations (i.e., rapid changes in sign) $ \chi(l) \le 1$. Moreover, if the function $f({\bf x})$ is self-similar, it can be expected that
\begin{equation}
\chi(l) \sim l^{-\kappa} , 
\label{eq:kappa}
\end{equation}
where $\kappa$ is the cancellation exponent. The exponent is a measure of the efficiency in the sign cancellations in the partition function, and if it exists (with $\kappa>0$), the function $f({\bf x})$ is said to be sign singular, as faster and faster changes in sign can be expected in the limit of infinite Reynolds number as smaller scales are considered. For a smooth function $\kappa=0$, while for a purely random process $\kappa=d/2$ where $d$ is the dimensionality of the system \cite{Ott92}.

\begin{figure}
\includegraphics[width=8cm,height=5cm]{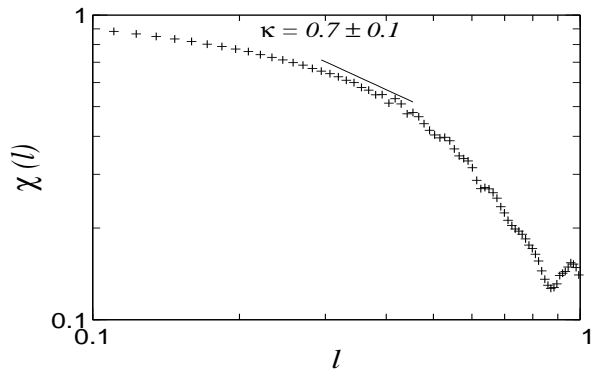} 
\caption{Partition function $\chi(l)$ for $u_x$ in run A1 (negligible rotation, ABC helical forcing). The slope corresponding to the cancellation exponent is indicated as a reference, in a range of scales that lies within the inertial range of the energy spectrum.}
\label{fig:chi_v_iso}
\end{figure}

It is interesting to point out that the cancellation exponent is related with fractal and conformal properties of a scalar distribution. In particular, it has been shown theoretically for a one-dimensional flow \cite{Vainshtein94}, and phenomenologically for 2D and 3D flows \cite{Sorriso02}, that the cancellation exponent can be related with the fractal dimension of structures. In particular \cite{Sorriso02},
\begin{equation}
\kappa = \frac{d-D}{2} - h ,
\label{eq:relation1}
\end{equation}
where $d$ is the space dimensionality, $D$ is the fractal dimension, and $h$ is the H\"older exponent of $f({\bf x})$ (i.e., the scaling exponent of its first order structure function).

When some properties of the flow are conformal invariant, another phenomenological relation between the cancellation exponent and Brownian diffusivity in conformal invariant processes belonging to a class of Schramm-L\"owner evolution (SLE) was obtained in \cite{Thalabard11},
\begin{equation}
1+\frac{\kappa_\textrm{SLE}}{8} = d-2{\kappa} ,
\label{eq:relation2}
\end{equation}
where $\kappa_\textrm{SLE}$ is the Brownian diffusivity of the SLE process (see \cite{Bernard06}), and is a number that characterizes to what class of universality the conformal process belongs. For the particular case of rotating helical turbulence, in \cite{Thalabard11} it was found that only the vertically-averaged vertical velocity and vorticity display conformal invariant behavior, with the same scaling for both the vertical velocity and vertical vorticity, and with $\kappa_\textrm{SLE} = 3.6 \pm 0.1$ measured for the vertical vorticity. It is important to note that the relation given by Eq.~(\ref{eq:relation2}) can only be expected to hold for quantities and systems that are conformal invariant.

\begin{table} 
\caption{\label{table:kappavel}Cancellation exponents in all the runs, for the three Cartesian components of the velocity.}
\begin{ruledtabular}
\begin{tabular}{ccccc}
Run/$\kappa$ & $u_x$ &     $u_y$    &     $u_z$     \\
\hline
T1 &   $0.7\pm 0.1$  &$0.7\pm 0.1$  &$0.9\pm 0.1$   \\
T2 &   $0.7\pm 0.1$  &$0.7\pm 0.1$  &$0.8\pm 0.1$   \\ 
T3 &   ---           & ---          &$0.7\pm 0.1$   \\ 
\hline
A1 &   $0.7\pm 0.1$  &$0.7\pm 0.1$  &$0.7\pm 0.1$   \\
A2 &   ---           & ---          &$0.35\pm 0.04$ \\ 
A3 &   ---           & ---          &$0.31\pm 0.02$ \\
\end{tabular}
\end{ruledtabular}
\end{table}

\section{Parallel computation}

The partition function $\chi(l)$ was computed for three-dimensional quantities and for vertically-averaged quantities (i.e., two-dimensional quantities). Given the spatial resolution of some of the simulations, a parallel method had to be developed to compute the cancellation exponent in three dimensions.

In practice, to compute the cancellation exponent in three dimensions, the cubic box of side $L=2\pi$ and volume $Q(L)$ gridded by $N^3$ points is divided into subvolumes without overlap, each with volume $Q_i(l)$ (where $l$ is the side of the subvolume), such that they cover the entire cubic box. For a given $l$, in each box the signed measure $\mu_i(l)$ is computed, and then the partition function $\chi(l)$ is built by summing over all $\mu_i(l)$. This process is repeated for all possible values of $l$. The smallest value of $l$ corresponds to the grid resolution, $l_\textrm{min}=2 \pi/N$, while the largest corresponds to the box side, $l_\textrm{max}=L$. For some values of $l$ (specially those close to $L$), the entire volume cannot be covered as the box is gridded by an integer number of points. In those cases, the method described in \cite{Graham05,Imazio10} was used to correct the normalization in Eq.~(\ref{eq:mu}). However, this results in fluctuations in $\chi(l)$ at large scales associated with finite box effects. The procedure for two-dimensional quantities is the same except for the change in dimensionality.

The computation of the cancellation exponent is easy to parallelize using the Message Passing Interface (MPI) library. The data was distributed among $N_p$ processors using the so-called 1D domain decomposition (see, e.g., \cite{Mininni11}), resulting in blocks of $N\times N\times (N/N_p)$ points in each processor. For sufficiently small subvolumes, $\mu_i(l)$ can be computed locally in each processor, and computation of $\chi(l)$ only requires a collective reduction to sum over all subvolumes in the different processors. For larger subvolumes (with side such that a subvolume spans several blocks of $N/N_p$ points in the vertical direction), extra communication is needed to compute $\mu_i(l)$ as the data in each subvolume may be distributed among several processors. A collective reduction is needed to compute $\mu_i(l)$ for each subvolume, and then another collective reduction is used to compute $\chi(l)$.

\begin{table} 
\caption{\label{table:kappaw}Cancellation exponents in the runs, for the three Cartesian components of the vorticity.}
\begin{ruledtabular}
\begin{tabular}{ccccc}
Run/$\kappa$ & $\omega_x$ & $\omega_y$ & $\omega_z$ \\
\hline
T1 &   $1.2\pm 0.1$  &$1.2\pm 0.1$  &$0.9\pm 0.1$   \\
T2 &   $1.2\pm 0.1$  &$1.2\pm 0.1$  &$0.9\pm 0.1$   \\ 
T3 &   ---           & ---          &$0.60\pm 0.03$ \\ 
\hline
A1 &   $0.9\pm 0.1$  &$1.0\pm 0.1$  &$1.0\pm 0.1$   \\
A2 &   ---           & ---          &$0.30\pm 0.07$ \\ 
A3 &   ---           & ---          &$0.29\pm 0.04$ \\ 
\end{tabular}
\end{ruledtabular}
\end{table}

\section{Results}

\subsection{Flows with negligible rotation}

The runs with negligible rotation (runs T1 and A1) have an isotropic energy spectrum compatible with Kolmogorov scaling (see \cite{Mininni09,Mininni09b}). In these runs, the same behavior was observed for the partition function $\chi(l)$ for the three Cartesian components of the velocity, as well as for the three Cartesian components of the vorticity. As an example, Fig.~\ref{fig:chi_v_iso} shows the partition function for $u_x$ in run A1. The partition functions for $u_y$ and $u_z$ look almost identical, as can be expected as for negligible rotation turbulence should be approximately isotropic. Note that for small values of $l$ the partition function approaches the asymptotical value of $1$, which corresponds to a smooth flow as can be expected at the smallest, dissipative scales (the value of 1 is indeed obtained for $l \approx 0.01$, the smallest scale available in the simulation, which is not shown in Fig.~\ref{fig:chi_v_iso} to zoom in inertial range scales). At the largest scales, the partition function is affected by the external forcing and by finite box size effects, and displays oscillations. We are interested in the intermediate range of scales that displays approximate power law behavior.

\begin{figure}
\includegraphics[width=8cm,height=5cm]{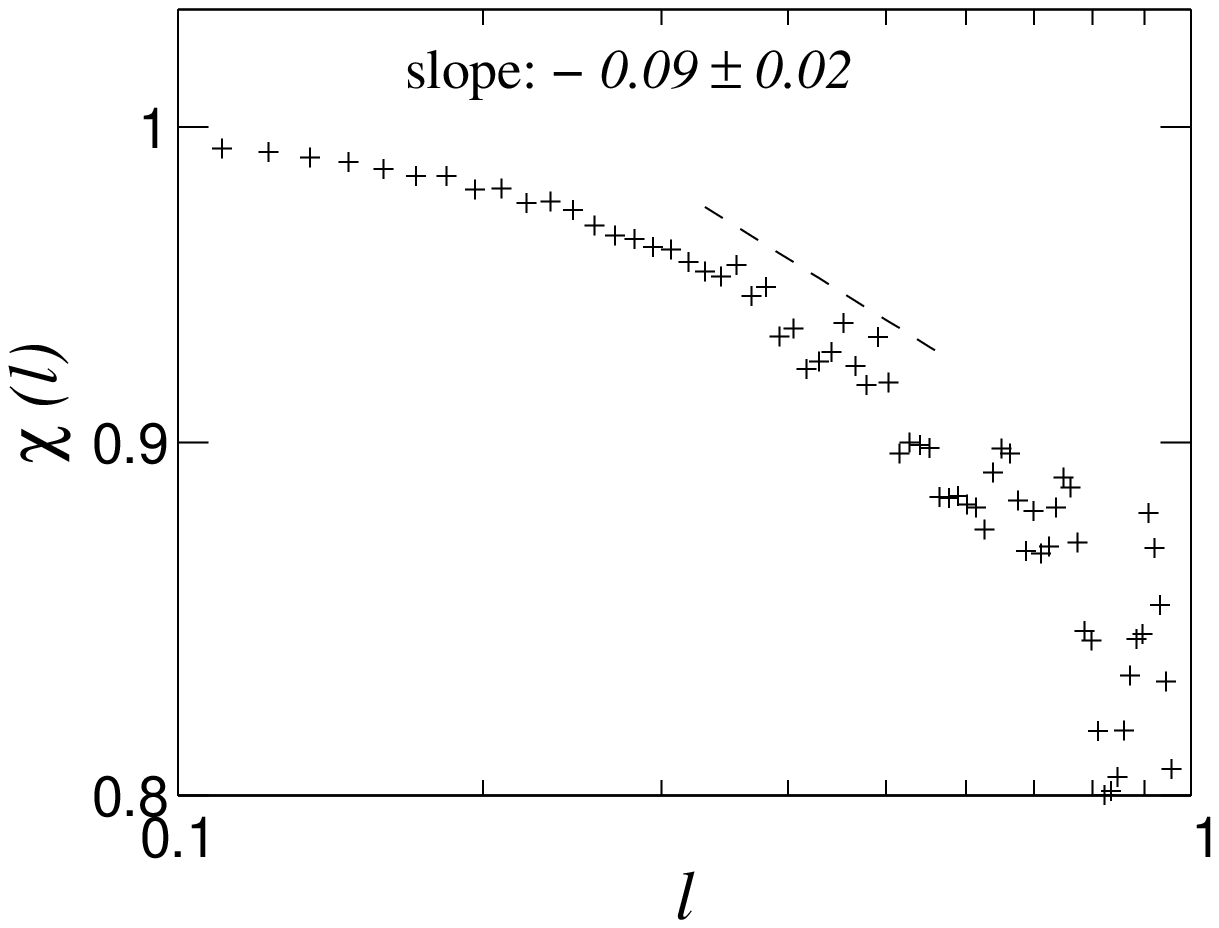} 
\includegraphics[width=8cm,height=5cm]{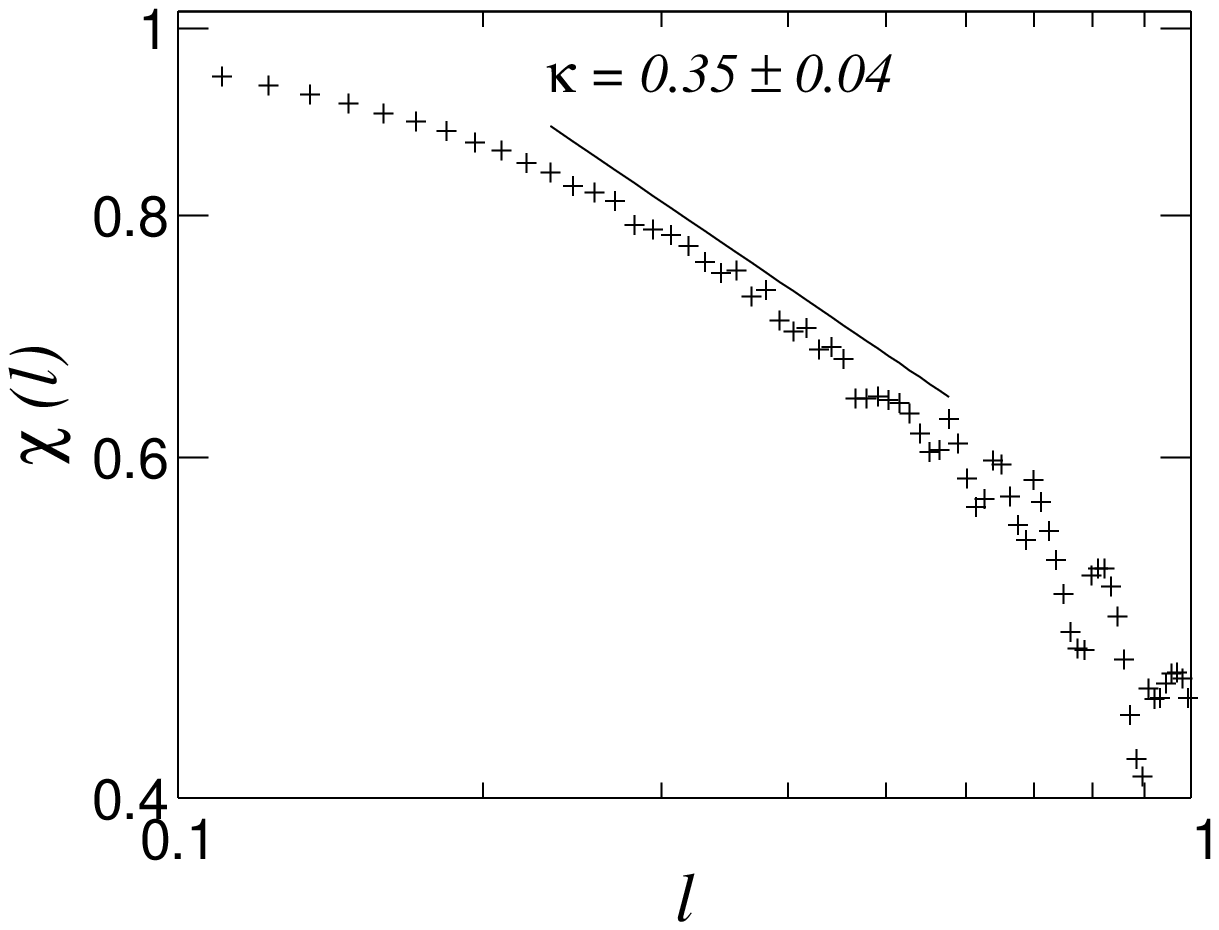} 
\caption{Partition function $\chi(l)$ for $u_x$ (above) and $u_z$ (below) in run A2 (with rotation, ABC helical forcing). The slope corresponding to the cancellation exponent is indicated by the solid straight line, in a range of scales that lies within the inertial range of the energy spectrum. Note that for $u_x$ no clear scaling is visible, and a slope (indicated by the dashed line) is only given as a reference to show the partition function is shallower than for $u_z$.}
\label{fig:chi_v_rot}
\end{figure}

\begin{figure}
\includegraphics[width=8cm,height=5cm]{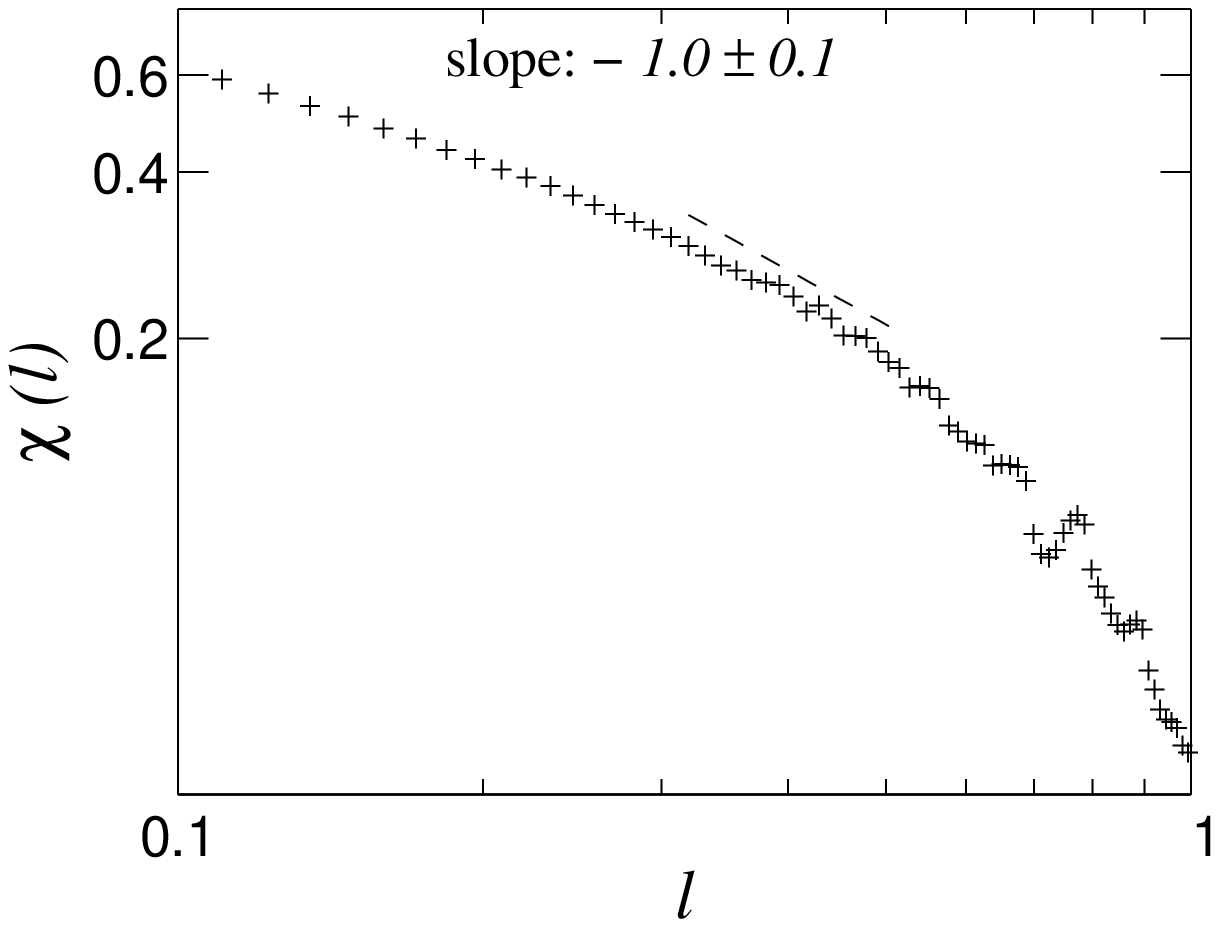} 
\includegraphics[width=8cm,height=5cm]{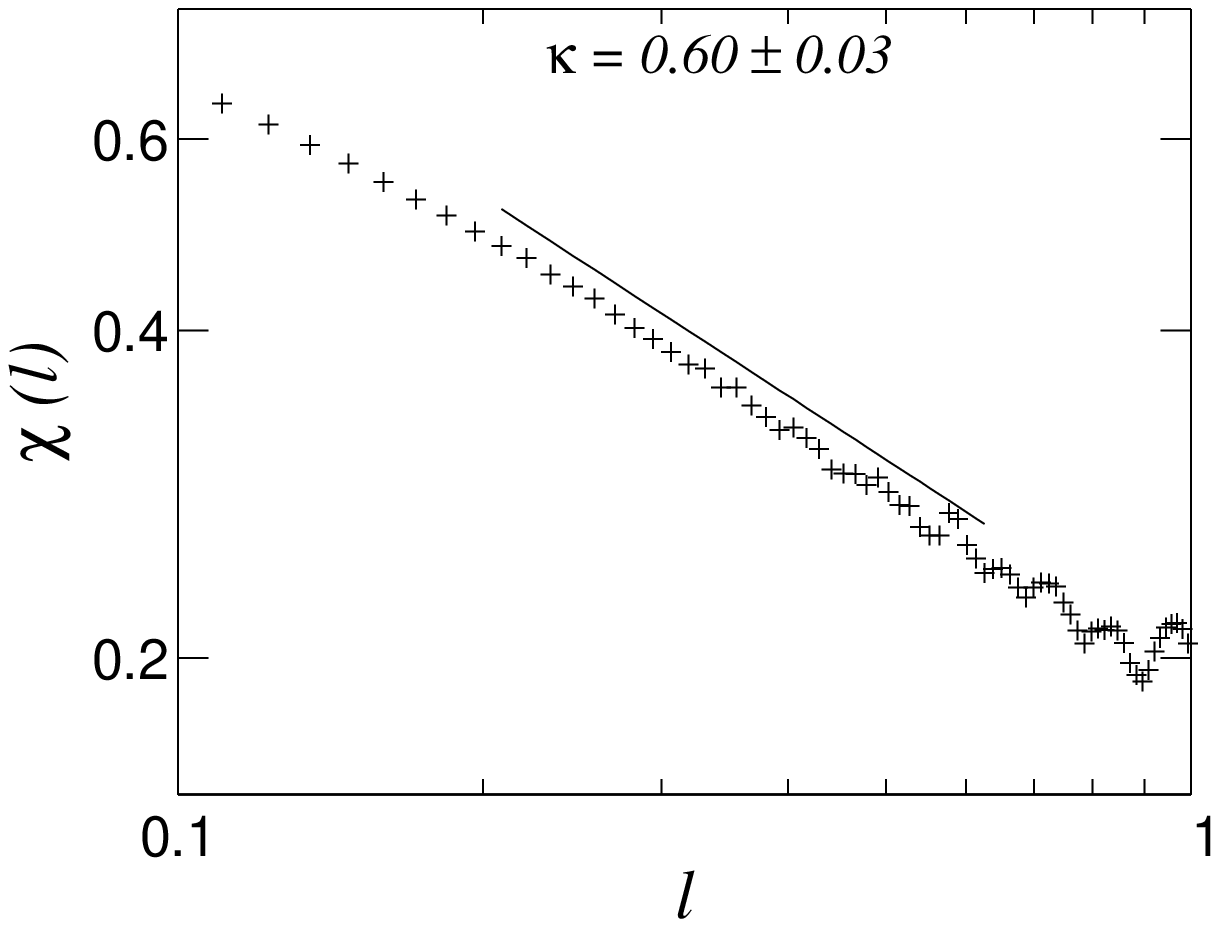} 
\caption{Partition function $\chi(l)$ for $\omega_x$ (above) and $\omega_z$ (below) in run T3 (with rotation, TG non-helical forcing). The slope corresponding to the cancellation exponent is indicated as a reference by the solid straight line, in a range of scales that lies within the inertial range of the energy spectrum. Note the shorter range of scales compatible with a power law in $\omega_x$ (with a slope indicated by the dashed line).}
\label{fig:chi_w_rot}
\end{figure}

Here and in the following, the cancellation exponent (or the absence of scaling behavior) is determined by looking at a range of scales that satisfies Eq.~(\ref{eq:kappa}), and by asking that the range should lie within (and be sufficiently wide when compared with) the inertial range identified in the energy spectrum of the same simulation. The cancellation exponent for the data in Fig.~\ref{fig:chi_v_iso}, obtained from a fit in this range of scales lying within the energy spectrum inertial range, is $\kappa = 0.7 \pm 0.1$. Note that this value is consistent with Eq.~(\ref{eq:relation1}), assuming that $h\approx 1/3$ (as expected for isotropic and homogeneous turbulence), and that the dimension of the structures in the flow is $D=1$ (i.e., vortex filaments).

The same cancellation exponent is obtained for all other components of the velocity field in run A1, and within error bars, also for the components of the velocity in run T1 (see Table \ref{table:kappavel}). However, $\kappa$ for $u_z$ in run T1 shows a slightly larger value than for $u_x$ and $u_y$. Although the difference is barely significative to indicate some weak anisotropy, this difference may be related with the fact that TG forcing only forces directly the $x$- and $y$-components of the velocity, while $u_z$ grows as a result of pressure fluctuations. Interestingly, run T2, with a Rossby number $Ro \approx 0.35$, still shows cancellation exponents for the velocity similar to those found in runs T1 and A1, indicating that smaller values of $Ro$ are required to observe the effect of the rotation in the velocity field.

The cancellation exponent can also be computed for the Cartesian components of the vorticity in these runs. Table \ref{table:kappaw} lists the values obtained for all the runs. In the runs with negligible rotation (see runs T1 and A1), similar values are still obtained in the three directions, but with larger differences between the values of $\kappa$ in the vertical and the horizontal components for the simulations that have no net helicity than in the case of the velocity.

\begin{figure}
\includegraphics[width=8cm,height=5cm]{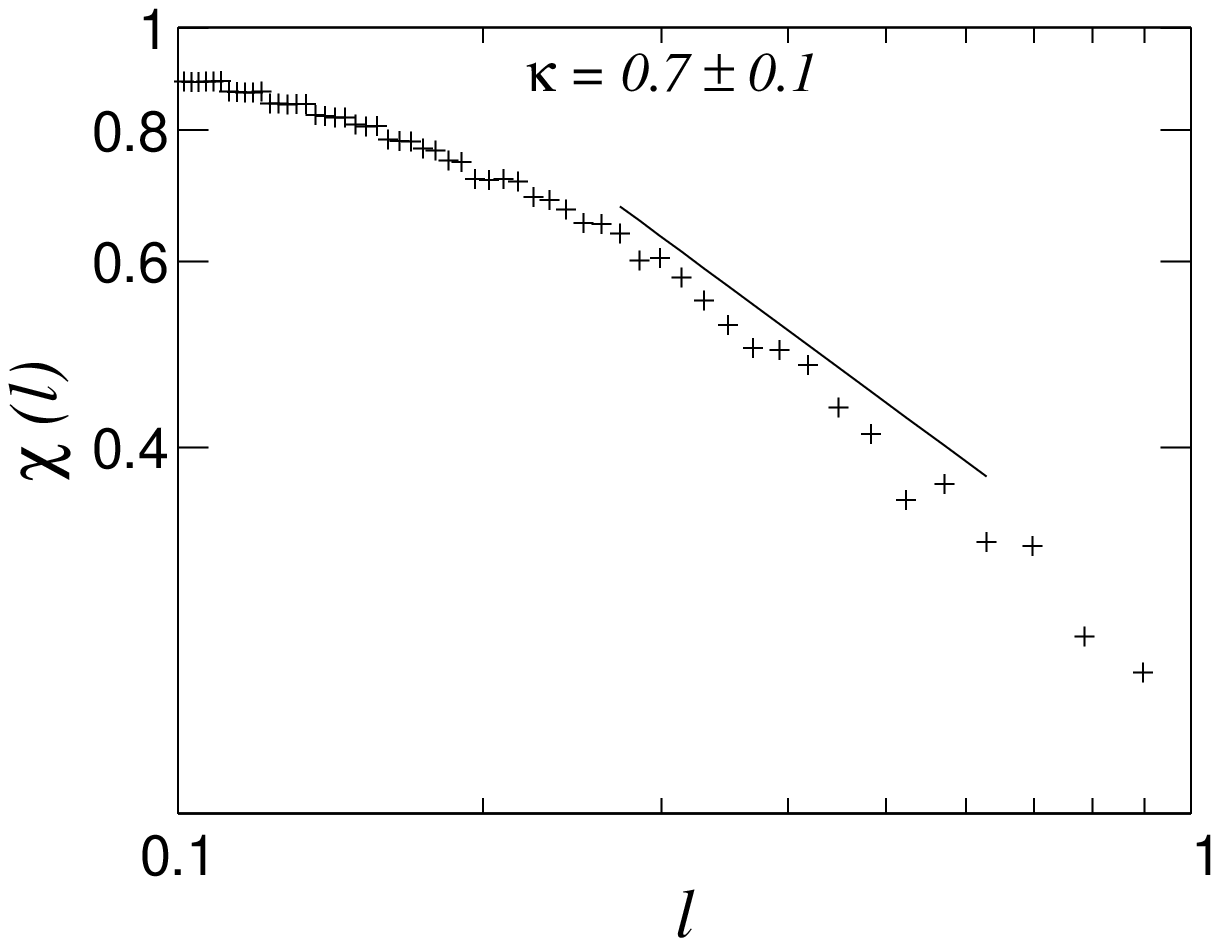} 
\includegraphics[width=8cm,height=5cm]{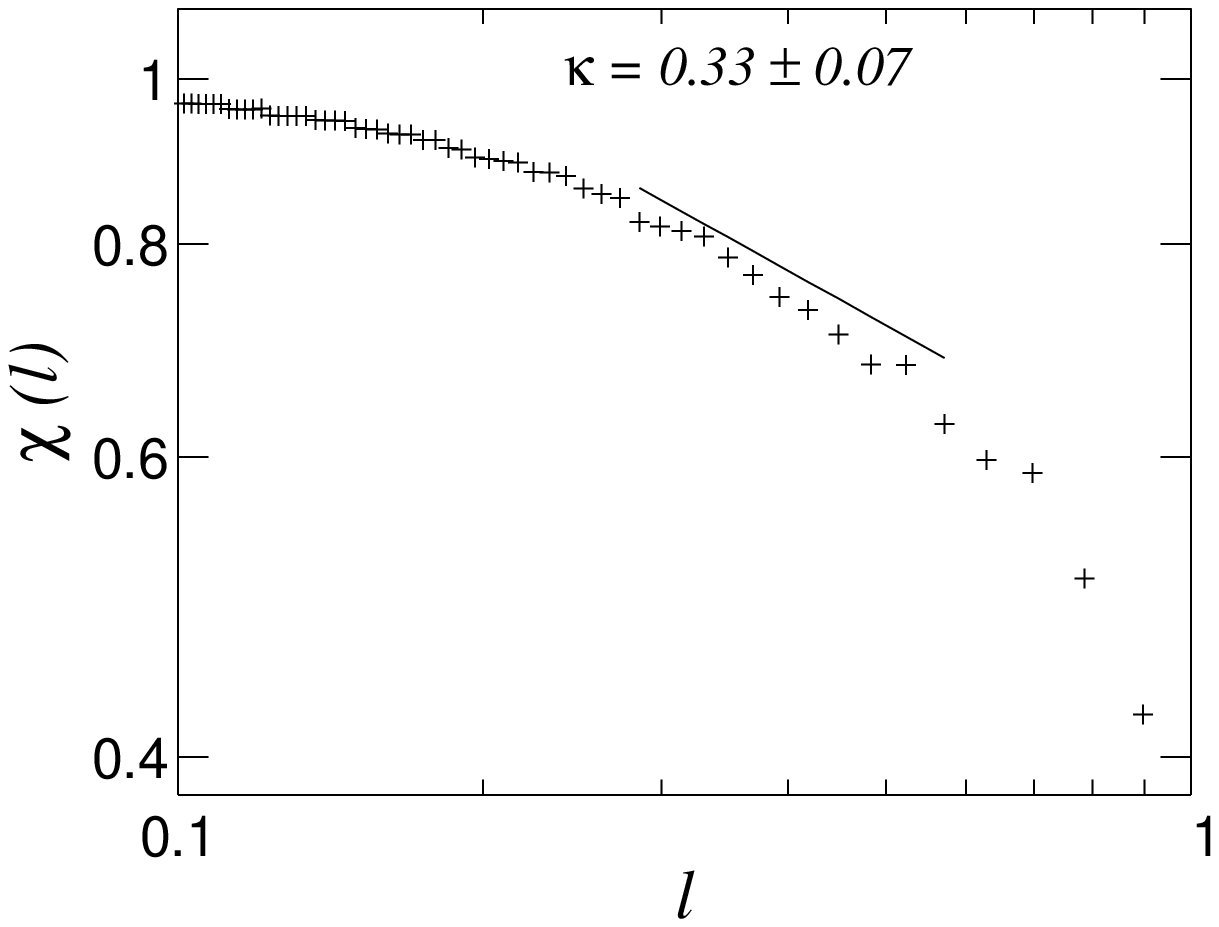} 
\includegraphics[width=8cm,height=5cm]{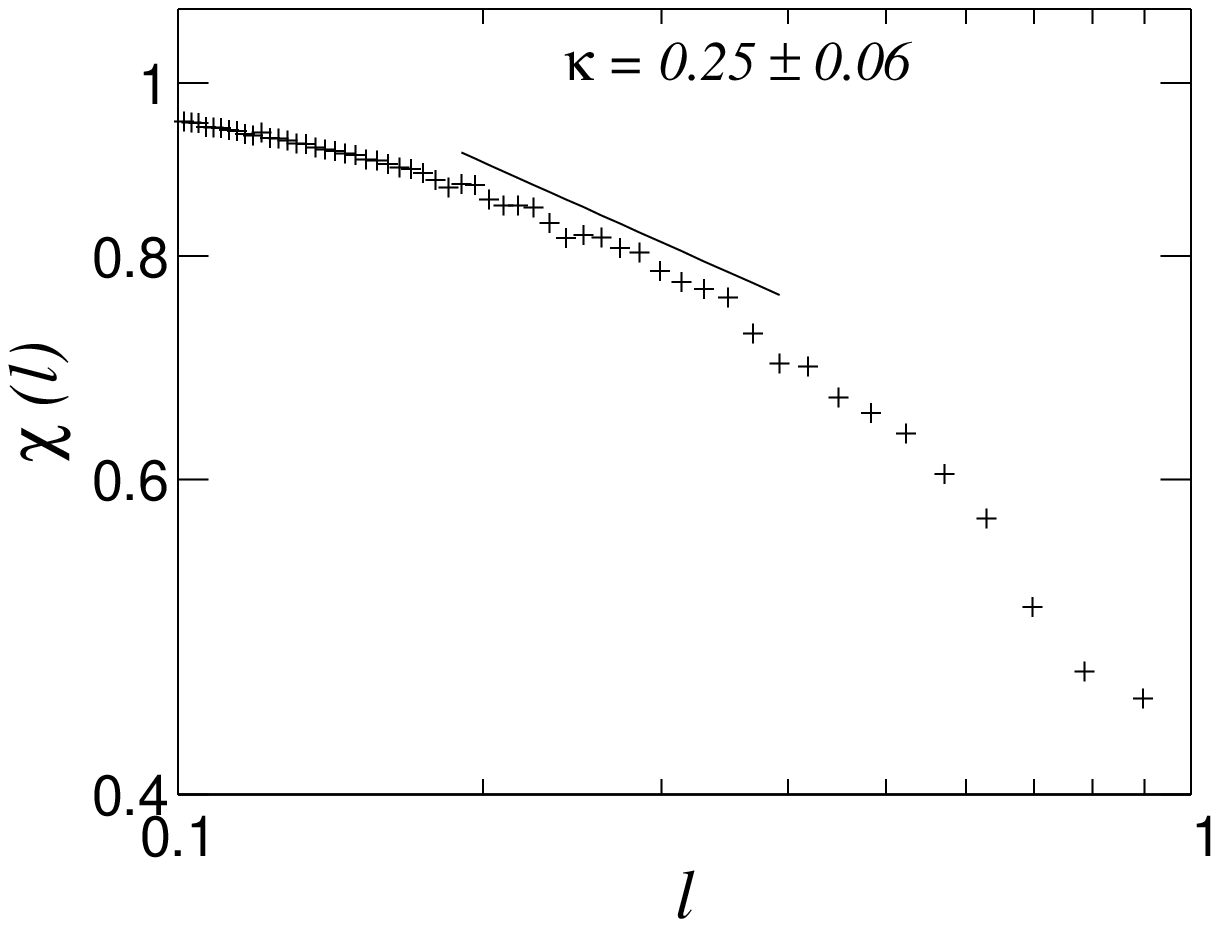} 
\caption{Partition function $\chi(l)$ for $\overline{u}_z$ in runs T3 (top), A2 (middle), and A3 (bottom). The slope corresponding to the cancellation exponent is indicated as a reference.}
\label{fig:chi_meanv}
\end{figure}

Finally, we computed the cancellation exponent for the helicity. As the cancellation exponent for the helicity in isotropic and homogeneous turbulence was studied in detail in \cite{Imazio10}, it suffices to say that the cancellation exponent obtained in these runs was consistent with that found in previous studies, with values $\kappa \approx 0.8 \pm 0.1$ for both runs, independently of whether the forcing is helical or not. This is consistent with the fact that in isotropic and homogeneous turbulence, helicity suffers a direct cascade with Kolmogorov scaling (see \cite{Chen03,Chen03b,Imazio10}). 

Note that with negligible rotation, the injection of helicity (in run A1) does not seem to significantly affect the scaling of $\kappa$ for any of the quantities studied (i.e., the scaling in the components of the velocity, the vorticity, or in the helicity itself, is similar to that found for run T1).

\subsection{Rotating flows}

We now consider the runs with smaller Rossby numbers. Rotation breaks isotropy and turbulence becomes anisotropic, with energy in spectral space transferred preferentially towards 2D modes, resulting in a quasi-bidimensionalization of the flow \cite{Cambon89}. The energy spectrum in the inertial range becomes steeper than in the isotropic and homogeneous case, as the result of the presence of waves which slow down the direct energy cascade \cite{Cambon97,Cambon04}. Figure \ref{fig:chi_v_rot} shows the partition function $\chi(l)$ in run A2 (with helical forcing) for the velocity components $u_x$ and $u_z$ (the behavior of $\chi(l)$ for $u_y$ is similar to that for $u_x$). While the partition function shows a clear range of scales with power law scaling for $u_z$, $\chi(l)$ is shallower for $u_x$ and shows large fluctuations, with almost no discernible inertial range. The cancellation exponent for $u_z$ in run A2 is $\kappa=0.35 \pm 0.04$, much smaller than in run A1 with negligible rotation (see Table \ref{table:kappavel}). A similar result was obtained in run A3, at much larger resolution (see Table \ref{table:kappavel}). In Table \ref{table:kappavel} we don't list values of $\kappa$ for $u_x$ and $u_y$ for runs in which no clear scaling can be observed in $\chi(l)$. However, if a power law fit is in any case attempted from the data in Fig.~\ref{fig:chi_v_rot} for $u_x$ in a range of scales corresponding to the inertial range, a slope $-0.09 \pm 0.02$ is obtained for a narrow range.

In run A2, the range of scales compatible with $\chi(l) \sim l^{-0.35}$ for $u_z$ indicates changes in sign in the vertical component of the velocity become more rapid as smaller scales are considered. In other words, the value of $\kappa$ is an indication that the vertical velocity in a rotating flow can be sign singular in the limit of infinite Reynolds number. On the other hand, the horizontal components of the velocity result in a much shallower distribution function, with slope closer to zero, an indication that these components are smoother and not sign singular.

\begin{figure}
\includegraphics[width=8cm,height=5cm]{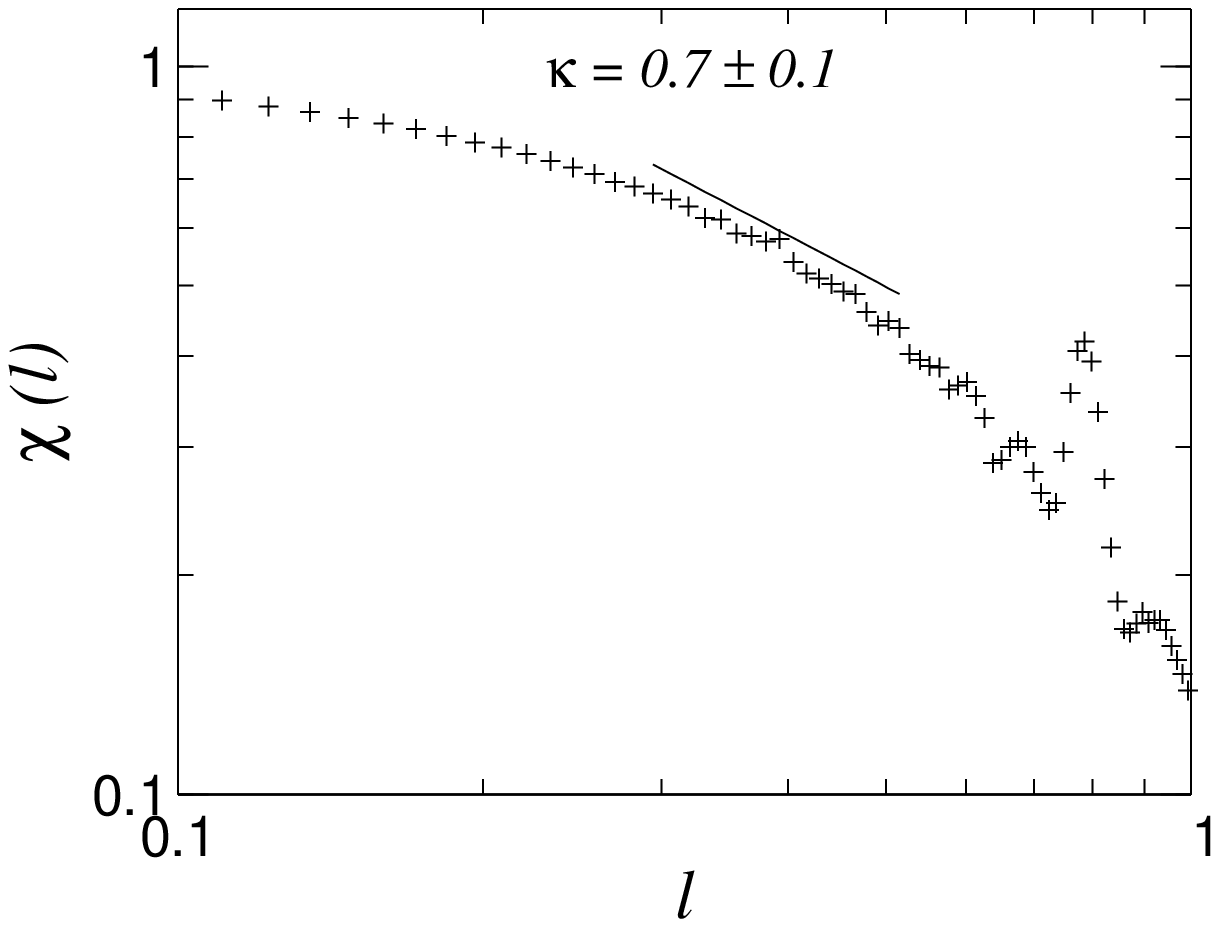} 
\includegraphics[width=8cm,height=5cm]{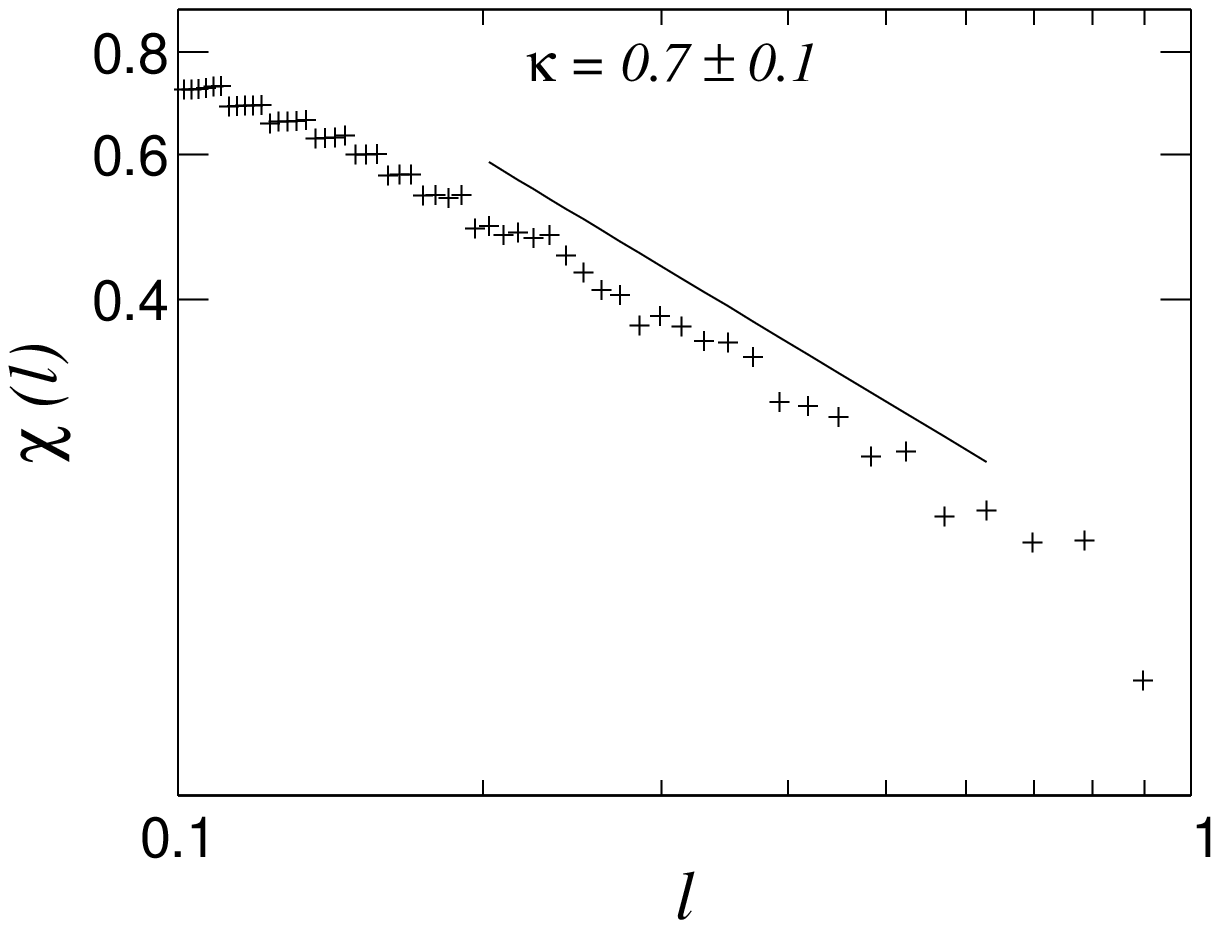} 
\includegraphics[width=8cm,height=5cm]{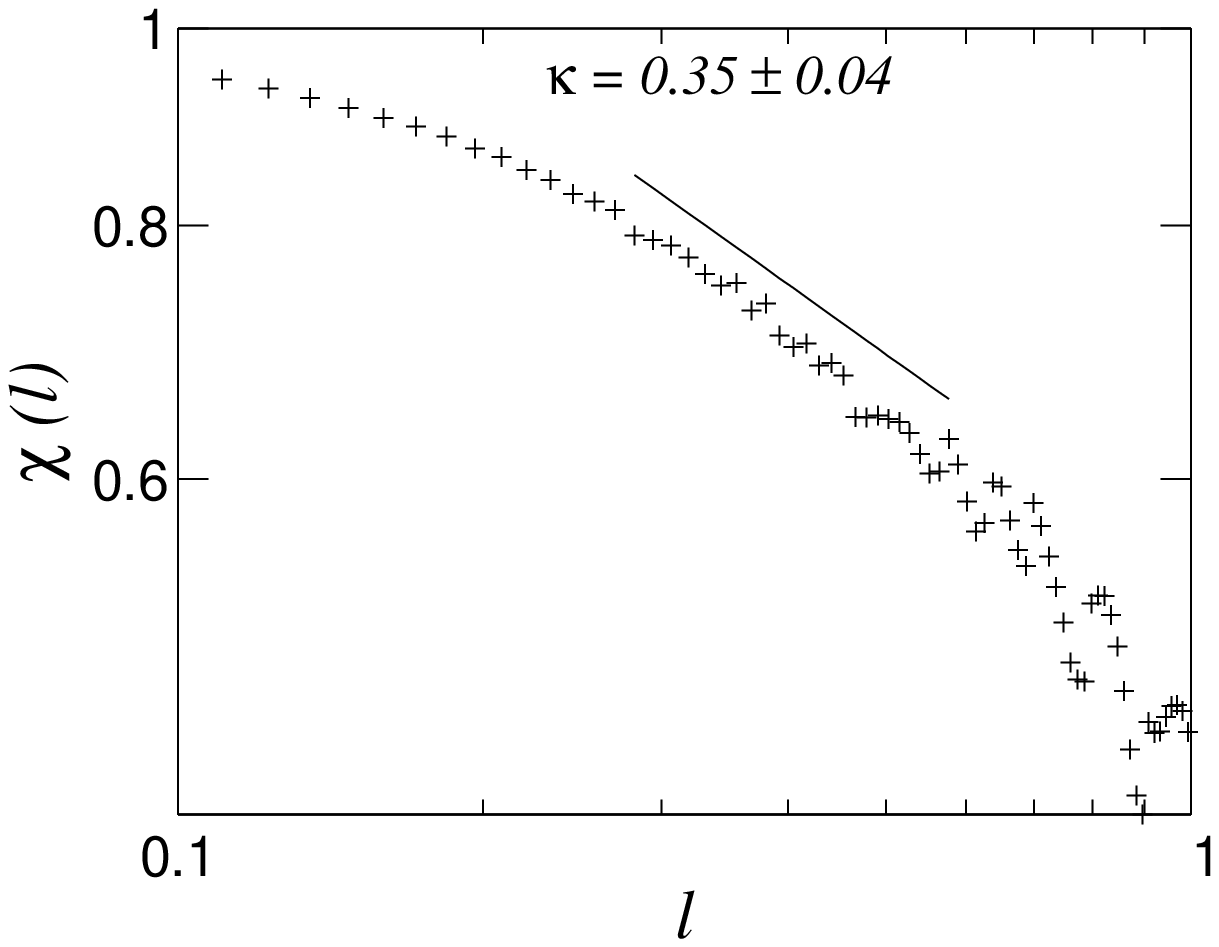} 
\includegraphics[width=8cm,height=5cm]{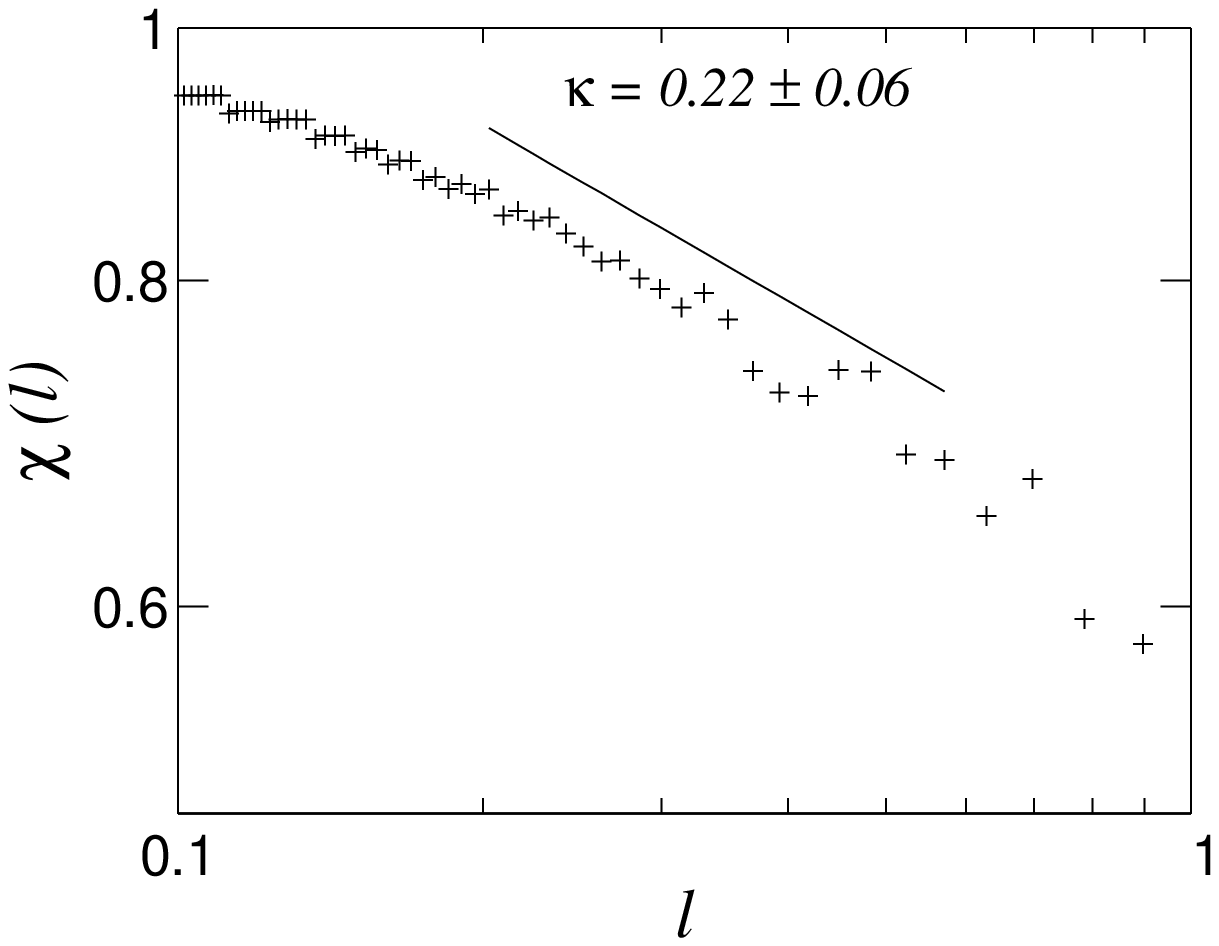} 
\caption{Partition function $\chi(l)$ for $h$ and $\overline{h}$. From top to bottom, partition function for: $h$ in run T3, $\overline{h}$ in run T3, $h$ in run A2, and $\overline{h}$ in run A2. The slope corresponding to the cancellation exponent is indicated as a reference.}
\label{fig:chi_helicity}
\end{figure}

Similar results are obtained for the runs with non-helical forcing. In run T3, $\chi(l)$ displays no scaling law for $u_x$ and $u_y$, while it shows a clear power law when computed for $u_z$ (see Table \ref{table:kappavel}). For $u_z$, the cancellation exponent resulted $\kappa = 0.7 \pm 0.1$, again showing that the vertical velocity in rotating flows is rough, with fast fluctuations, and may be sign singular in the limit of infinite Reynolds number. However, note that unlike the case of negligible rotation, the runs with and without helicity show different scaling laws for $u_z$ (compare $\kappa=0.35 \pm 0.04$ in run A2 with $\kappa = 0.7 \pm 0.1$ in run T2). This difference can be understood as helical and non-helical rotating flows are known to follow different scaling laws (i.e., they have different H\"older exponents, see \cite{Mininni10}), and as passive scalars advected by those flows also have different scaling as a result \cite{Imazio11}.

The cancellation exponent for the Cartesian components of the vorticity shows a similar behavior. In all runs with small enough Rossby number, the partition function $\chi(l)$ is shallower for $\omega_x$ and $\omega_y$, and steeper for $\omega_z$. As for the velocity, a clear range with power law behavior can be identified for $\omega_z$, while for $\omega_x$ and $\omega_y$ the range of scales compatible with a power law in the partition function is either significantly narrower than the inertial range, or inexistent. As an example, Fig.~\ref{fig:chi_w_rot} shows the partition functions for $\omega_x$ and $\omega_z$ in run T3. The resulting cancellation exponent is indicated for $\omega_z$, while for $\omega_x$ a slope for scales in the inertial range is only given as a reference. Table \ref{table:kappaw} lists the values of $\kappa$ obtained in all the runs.

From the values in Tables \ref{table:kappavel} and \ref{table:kappaw}, an estimation of the fractal dimension of structures can be obtained from Eq.~(\ref{eq:relation1}). As an example, for run T3, and using the value of $h$ obtained from numerical simulations of non-helical rotating turbulence ($h \approx 1/2$, see \cite{Muller07,Mininni09}), the fractal dimension obtained is $D = 0.6 \pm 0.2$ for $u_z$ and $D = 0.8 \pm 0.2$ for $\omega_z$. Similar values are obtained for runs A2 and A3. The values, close to one, are compatible with column-shaped structures, as are often observed in rotating flows.

Leaving aside the fractal dimension, several things are worth pointing out from the values of the cancellation exponent obtained. As in the case of the vertical velocity, the vertical vorticity shows signs of being sign singular in the limit of infinite Reynolds number. Moreover, $u_z$ and $\omega_z$ have (within error bars) the same cancellation exponent in the runs with helical forcing, as well as in the runs with non-helical forcing (although the actual value of $\kappa$ depends on whether the forcing is helical or not, as discussed above). This is consistent with the prediction that to the lowest order in an expansion with the Rossby number as small parameter, the vertical components of the velocity and vorticity satisfy the same equation, namely that of a passive scalar in two-dimensions (see Eqs.~\ref{eq:vz} and \ref{eq:wz}).

However, strictly speaking Eqs.~(\ref{eq:vz}) and (\ref{eq:wz}) apply to the vertically averaged fields $\overline{u}_z$ and $\overline{\omega}_z$. Figure \ref{fig:chi_meanv} shows the partition functions and cancellation exponents $\kappa$ obtained for $\overline{u}_z$ in runs T3, A2, and A3. The cancellation exponents for $\overline{\omega}_z$ are $\kappa = 0.31 \pm 0.04$ in run T3, and $\kappa = 0.30 \pm 0.07$ in run A2 (with a similar value in run A3). Except for the vertical vorticity in run T3 (for reasons that may be related with properties of the TG forcing as explained above), the cancellation exponents for the 3D quantities and for the vertically averaged quantities give similar values.

In the helical rotating case, $\kappa \approx 0.3$ for both $\overline{u}_z$ and $\overline{\omega}_z$. As mentioned in the introduction, isocontours of these quantities are known to correspond to conformal invariant SLE behavior with associated diffusivity $\kappa_\textrm{SLE} = 3.6 \pm  0.1$. SLE behavior has been found in the past for quantities that are advected as active scalars by a self-similar flow \cite{Bernard07} (note that although Eqs.~\ref{eq:vz} and \ref{eq:wz} indicate the quantities behave as passive scalars, at finite Rossby numbers these quantities affect the flow evolution). As the vertically averaged quantities are defined in a space of dimensionality $d=2$, from Eq.~(\ref{eq:relation2}) it follows that for $\kappa \approx 0.3$ then $\kappa_\textrm{SLE} \approx 3.2$, close to the value obtained independently from an analysis of conformal invariance using the same dataset \cite{Thalabard11}. As a result, these independent measurements of the cancellation exponent confirm the validity of the relation between $\kappa$ and $\kappa_\textrm{SLE}$ obtained in \cite{Thalabard11} for SLE systems.

Finally, we also computed the cancellation exponent for the helicity in the runs with non-negligible rotation. As in isotropic and homogeneous flows, the helicity is an invariant in the absence of dissipation. As a result, in the viscous case it suffers a direct cascade, and this results in a power law in its inertial range spectrum \cite{Mininni09b}. The analysis of the helicity was done for the 3D quantity as well as for the vertically-averaged quantity, and the results are shown in Fig.~\ref{fig:chi_helicity}. The partition functions show clear scaling laws, with a smaller cancellation exponent in the runs with helical forcing than in the runs with non-helical forcing. This in in agreement with previous studies that showed that the scaling laws in rotating turbulent flows are affected by the presence of helicity (see \cite{Mininni09b}).

\section{Conclusions}

We computed the cancellation exponent for several quantities in direct numerical simulations of rotating turbulent flows with different forcing functions. The exponent allowed us to study the statistics of fast fluctuations and sign cancellations in quantities of interest at different scales. The flows were forced with two different forcing mechanisms: an Arn'old-Beltrami-Childress (ABC) maximally helical forcing, and  a Taylor-Green (TG) non-helical forcing. The simulations analyzed had spatial resolutions ranging from $512^3$ to $1536^3$ grid points, Reynolds numbers between $\approx 1100$ and $\approx 5100$, and Rossby numbers between $\approx 0.06$ and $\approx 8$.

The cancellation exponent was computed for the Cartesian components of the velocity field and of the vorticity, for the helicity, and for vertically-averaged quantities. In the runs with negligible rotation, the cancellation exponent is the same in the three directions for the velocity and the vorticity, as expected from the isotropy of the flow. The exponents obtained are also consistent with scaling laws expected for isotropic and homogeneous turbulence. Finally, the cancellation exponent found for the helicity is consistent with previous studies and with Kolmogorov scaling.

In the runs with rotation, there is a large difference between the behavior of the $x$- and $y$-components of the velocity and vorticity fields, and the $z$-components. While the horizontal components have a shallower partition function (specially for the velocity, for which the partition function is almost flat), the vertical components show clear scaling laws and the behavior is compatible with a rough field that develops faster fluctuations and faster changes in sign at smaller scales (i.e., the behavior is compatible with sign singularity in the limit of infinite Reynolds number). Considering the symmetries of rotating turbulence, the analysis was extended also to vertically averaged quantities, confirming the results. The values found are consistent with the scaling laws known to be followed by the energy spectrum in rotating turbulence.

Moreover, the values of the cancellation exponent found for the vertical components of the fields in the case of rotating helical turbulence are in agreement with what can be expected from previous results that found that these quantities display SLE behavior. The values of the cancellation exponent found in this study confirm a phenomenological relation for conformal invariant systems between the cancellation exponent and the diffusivity of the Brownian process in the equivalent SLE system derived in \cite{Thalabard11}.

The facts that $u_z$ and $\omega_z$ show similar scaling laws, and that the values found for the cancellation exponents in the helical case satisfy the phenomenological relation expected for SLE processes (which is expected to hold for active scalar quantities that are advected by a rough self-similar field), are consistent with theories that predict that to the lowest order in an expansion in terms of the Rossby number, these two quantities satisfy the same dynamical equation. While $u_z$ and $\omega_z$ are rough, the horizontal components of the field are observed to be smooth and to display less fluctuations.

Overall, the cancellation exponent results in a useful tool to characterize scaling laws in rotating turbulent flows, allowing to discriminate between different field components. Together with the relations discussed in this paper, it can also give a method to estimate scaling laws from direct measurements in experiments, e.g., of only one component of the velocity, even when averaged along one direction \cite{Cobelli09,Cobelli09b}.

\begin{acknowledgments}
The authors acknowledge support from grants No.~PIP 11220090100825, UBACYT 20020110200359, and PICT 2011-1529 and 
2011-1626. PDM acknowledges support from the Carrera del Investigador Cient\'{\i}fico of CONICET. 
\end{acknowledgments}

\bibliography{ms}

\end{document}